\documentstyle[editedvolume,amsmath,numreferences,epsfig]{crckapb} 

\begin{opening}
\title{Full Counting Statistics in electric circuits}

\author{Markus Kindermann}
\institute{Instituut-Lorentz, Universiteit Leiden, P.O. Box 9506, 2300 RA
Leiden, The Netherlands}
\author{Yuli V. Nazarov}
\institute{Department of Nanoscience, Delft University of Technology,
Lorentzweg 1, 2628 CJ Delft, The Netherlands}
\end{opening}
\begin{document}

\section{Introduction}	

Full counting statistics (FCS) is one of the most attractive
and intellectually involved concepts in quantum transport. 
It provides much  information about charge transfer (all 
moments, including average current, noise, etc.) in a compact and elegant
form. Albeit the study of full counting statistics has begun with 
some confusion. The first attempt to derive FCS \cite{Lev92} exploited 
the text-book definition of quantum measurement: the probability of the 
outcome $q$ of a measurement is given by $\langle n|\delta(q -\hat Q)|n\rangle$, $\hat Q$
being an operator associated with the value measured and  $|n\rangle $ the quantum state.
However, the choice of  $Q$ made in  \cite{Lev92}
resulted in severe interpretation problems.
In \cite{Lev93} the authors revised their approach.  
 The new method, that is  commonly accepted now, invokes an extra degree of freedom,
a detector. The quantum measurement paradigm is applied to the detector degree of
freedom.

There is a similarity between the method of \cite{Lev93} 
and two core approaches to quantum dynamics: the  Keldysh technique \cite{Rammer} and the 
Feynman-Vernon formalism \cite{Fey63}. This similarity was not stressed in \cite{Lev93}.
Later it has been noticed that the FCS can be evaluated with a straightforward 
modification of the  Keldysh Green function technique \cite{Naz99}. This allows one to extend
the studies of the  FCS to  many systems, 
some results being reviewed in other contributions to this  book.
This is  good news. Seemingly bad news is that the method of \cite{Lev93}
does not always give results that can be interpreted as probabilities
of  measurement outcomes. In the statistics of charge transfer,
the bad news becomes relevant for systems where gauge invariance is broken 
\cite{Bel01,She02}. If one tries to generalize FCS to 
arbitrary variables, these problems arise from the very beginning \cite{Naz01}.

The present contribution consists of two parts. First we address
the bad news. For this, we consider a rather general and abstract problem, the  counting statistics of a general quantum mechanical variable being measured
by a linear detector. We find that in this case the FCS is physically meaningful 
and useful since it determines the quantum time evolution of the detector.
This is despite the fact that it can not be interpreted as a probability distribution \cite{Naz01}.

Thus encouraged, we investigate how far one can go with this approach.
The  abstract exercise with a linear
detector proves to be very useful to describe a mesoscopic 
conductor embedded in  a linear electric circuit.   
We reveal the relations between the FCS and  the non-equilibrium Keldysh
action that describes current and voltage fluctuations
in electric circuits. We show how to evaluate 
the current and voltage counting statistics at any two contacts
of an arbitrary circuit. This appears to be relevant for
future experimental activities in the field of 
quantum noise and statistics \cite{Kin02}.

The article is organized as follows. 
In section \ref{sc:elementary} 
we will pursue a  route similar to that proposed in \cite{Lev92}, 
defining a current statistics as the statistics of a charge operator. This route has not been explored before. It  allows to understand the origin of the ``negative probabilities'' found in \cite{Bel01} purely in terms of conductor degrees of freedom. 
We follow an  alternative route in section \ref{sc:static} 
and consider a system coupled to  a static linear detector \cite{Naz01}.
We show that the  result  of the detection depends 
in  general  on the initial state of the detector. 
To predict the result, one needs a function of two variables. 
We adopt then this function as the {\it definition}
of the  full counting statistics.   
We will identify situations where the two  approaches
give the same easy-to-interpret result.

Then we turn to electric circuits. We explain in section \ref{sc:linear}
why a linear circuit can be viewed as a set of
dynamical linear detectors 
and why the FCS expression is the building block of the 
Keldysh action that describes the quantum dynamics of the circuit. 
In section \ref{sc:low} we discuss the {\it low-frequency} 
limit of the action. In this limit, the action is local in time. This
facilitates its evaluation.
The general scheme is illustrated in  section \ref{sc:QPC} with 
a simple model of a phase-coherent conductor in series with an external
linear resistor. In this case, we are able 
to determine the full statistics of current
and voltage fluctuations.  These results were obtained in collaboration with
C.W.J. Beenakker, that we gladly acknowledge.

\section{Charge statistics without detector} \label{sc:elementary}

We start out by defining a statistics of time averages of electric
current  without specifying a measurement
procedure, as it has been also done in \cite{She02}. Unlike in \cite{She02} we  study directly the statistics of charge $Q$
on one side of the cross-section of a conductor  through which the current is measured.
If $Q$ is fixed at time $0$, the statistics of $Q(\tau)$ at
time $\tau$ corresponds to the statistics of the charge that has traversed the cross-section in 
the time interval $[0,\tau]$. 
Having the clear physical interpretation of the statistics of charge on one side of the cross-section, this definition will shed some light on interpretation problems encountered with other definitions, like "negative probabilities" and half-integer charge transfer \cite{Bel01,She02}.

To specify $Q$ we introduce, following \cite{Lev93}, a smooth function
$f$ that divides the conductor into two parts: a left side, where
$f({\bf x})<0$, and a right side, where $f({\bf x})>0$. We are interested in charge transfer through the boundary $f({\bf x})=0$
between
these two sides. The operator of electric
charge to the right of the cross-section is  $Q=e\theta[f({\bf
x})]$ [where $\theta(x)=0$ for $x \leq 0$ and $\theta(x) =1$ for $x>0$], $e$
being the elementary charge. It is convenient to express the statistics
of $Q(\tau)$ via the corresponding
 moment generating function \begin{equation}
\label{eq:gendef} \chi_{\rm c}(\lambda) =\sum_k {\frac{(i\lambda)^k}{k!}
\left\langle Q(\tau)^k \right\rangle} = \left\langle e^{i\lambda
Q(\tau)} \right\rangle = \left\langle e^{i H \tau} e^ {i\lambda Q}
e^{-iH\tau} \right\rangle. \end{equation} 
Here $H$ is the
Hamiltonian of the current conductor and the average is taken over the
initial density matrix of the conductor.

One advantage of defining $\chi_{\rm c}$ as in  Eq.\ (\ref{eq:gendef})
is that it is evidently  associated with a (positive) probability
distribution,  the distribution of measurement outcomes of an observable  corresponding to the Hermitian operator
$Q(\tau)$. Besides, $\chi_{\rm c}$  predicts   charge 
transfer  in integer multiples of the electron charge $e$ for
systems of well-localized non-interacting electrons,
the result to be expected.
Both very physical properties have been found to be violated by other definitions
\cite{Bel01,She02}. One buys these advantages by defining a statistics that is
only indirectly linked to the statistics of transmitted charge. That link 
requires the
knowledge of the initial charge state of the conductor. 

Using that
 \begin{equation} e^{i\lambda Q/2} e^{A} e^{-i\lambda
Q/2}= \exp[e^{i\lambda Q/2}A e^{-i\lambda Q/2}] 
\end{equation}
 we rewrite Eq.\ (\ref{eq:gendef}) identically as \begin{eqnarray} \chi_{\rm
c}(\lambda) = \left\langle e^{i\lambda Q/2} \exp\left[i e^{-i\lambda
Q/2}H e^{i\lambda Q/2} \tau\right] \right. \nonumber \\ \left.
\exp\left[-i e^{i\lambda Q/2}H e^{-i\lambda Q/2} \tau\right] e^{i
\lambda Q/2} \right\rangle. \end{eqnarray} The charge operator $Q$
commutes with all position operators ${\bf x}$ contained in the
Hamiltonian $H=H({\bf p},{\bf x})$. It does, however, not commute with the
momentum operators ${\bf p}$.  As a consequence,   momentum operators are
transformed as 
\begin{equation} e^{-i\lambda Q/2} {\bf p} e^{i\lambda
Q/2}= {\bf p} - {\textstyle \frac{e}{2}} \lambda \boldsymbol{\nabla}
\theta[f({\bf x})] \equiv \tilde{\bf p}_{\lambda}. 
\end{equation} 
We  define a new Hamiltonian $H_{\lambda}= H(\tilde{\bf p}_{\lambda},{\bf
x})$ in the same way as it has been done in \cite{Lev93}. The generating
function 
\begin{equation} \label{eq:chic} 
\chi_{\rm c}(\lambda) =
\left\langle e^{i \lambda Q/2} e^{i H_{-\lambda} \tau} e^{-iH_{\lambda}
\tau} e^{i \lambda Q/2} \right\rangle \end{equation} takes then a form
that is very similar to the $\chi(\lambda)$ found in \cite{Lev93} within
the spin-$\frac{1}{2}$ current detection model . Tracing back the
difference we write $\chi(\lambda)$ in terms of charge operators as
\begin{equation} \label{eq:chifromchic} 
\chi(\lambda) = \left\langle
e^{-i\lambda Q/2} e^{i\lambda Q(\tau)} e^{-i\lambda Q/2} \right\rangle .
\end{equation} 
If the initial state of the conductor is an eigenstate of
charge with eigenvalue $Q_0$, the two generating functions are identical
up to the offset charge $Q_0$, $\chi(\lambda) = e^{-i\lambda Q_0}
\chi_{\rm c}(\lambda)$ . In this case both characteristic functions are
associated with a positive probability 
\begin{equation} P(q) =
\int{d\lambda \, e^{-i q \lambda} \chi(\lambda)} 
\end{equation} 
to
transfer $q$ charges during the measurement time. For generic systems of
non-interacting electrons this probability has been found to be non-zero
only at integer multiples of the electron charge, corresponding to the
transfer of an integral number of electrons \cite{Lev93}. 

If the initial state is a superposition of eigenstates of charge $Q$,
 the two generating functions $\chi$ and $\chi_{\rm
c}$ differ. For example, $\chi$ may seem to predict the transfer of half
the elementary charge for systems of well-localized non-interacting electrons, when $\chi_{\rm c}$  indicates integer charge transfer.   This
becomes evident when $\chi$ is written in the form
(\ref{eq:chifromchic}). It contains
summands of the form $\chi_{mn}=\exp[-i\lambda(m+n)/2] \left\langle
m,\alpha| \exp[i \lambda Q(\tau)]|n,\beta \right\rangle$, 
$|m,\alpha\rangle$ and $|n,\beta\rangle$ being eigenstates of the charge
with additional quantum numbers $\alpha$, $\beta$, $Q|m,\alpha\rangle =
m|m,\alpha\rangle$. While the matrix element in the expression for
$\chi_{mn}$ corresponds to integer charge transfer, the pre-factor
suggests the transfer of half-integral charges  when $m+n$ is an odd number. This
has been observed in \cite{She02}. 

For a Josephson junction at the cross-section $f({\bf x})=0$ between two
superconductors the situation is even more involved. One would like to
choose an initial state of well defined phase difference between the two
sides of the junction in order to have a well defined current flowing.
However,  phase and charge  are
conjugated variables obeying Heisenberg's uncertainty
principle and  the dispersion of the initial charge $Q$ in such a 
state is
infinite.  The
charge distribution corresponding to $\chi_{\rm c}$ with  an initial state of well-defined phase
is therefore one of undetermined $Q(\tau)$. It contains
no information about the
charge transfer. This is clearly undesirable. Because all the
uncertainty of $Q$ is already present in the initial state and has
nothing to do with the transfer process, one might hope to be able to
eliminate it in a meaningful way. Indeed, we have seen that this is
accomplished for the initial charge offset in a state of well defined
charge by the factors $e^{-i\lambda Q/2}$ in the definition of $\chi(\lambda)$
(\ref{eq:chifromchic}). One could hope that this works also for the
initial charge spread in a phase eigenstate. Instead, $\chi(\lambda)$ then seems to imply  ``negative probabilities''
 \cite{Bel01}. 

A way to remedy this problem would be to calculate $\chi_{\rm c}$ in an
initial state with a finite charge and phase dispersion. One could also
try to find general relations between the charge statistics and the
initial state of the conductor that would characterize the charge
transfer process more generally.

The alternative is to couple the
(super-)conductor to a detector and interpret the detector read-off in
terms of charge transfer. We will follow this route in the next
section. It turns out that in a idealized detector model a
function  very similar to the generating function $\chi$
eventually determines the final state of the detector
provided the initial state is known.
The charge transfer through the conductor
is then  characterized by its effect on the detector.

\section{Counting statistics with a static detector} \label{sc:static}

We turn now to the {\em measurement} process of time averaged
quantities. In this section we focus on an idealized measuring device  without
its own dynamics, a static detector. Within this model we study the measurement of time
averages $\int_0^{\tau} dt \, A(t)$ of an arbitrary operator $A$.
For electric currents, $A=I$, this will allow us to characterize
the charge transfer and its statistics by a function similar to the
characteristic function introduced in the previous section. The analysis
of this function will allow us to trace the origin of the "negative
probabilities" found in \cite{Bel01}. 

\subsection{Detector Model}
\label{model}

 We employ a detector model that has first  been used by John von Neumann  in an analysis  of the quantum measurement process. Following  \cite{Neumann}
e introduce a detector
variable  whose operator $ x$ commutes with all operators
of the system to be measured.
Its Hamiltonian is
  $ {q}^2/2m $, where  $q$ is the variable  conjugated to $x$.
The system measured shall be coupled to the position $x$ of the  detector during the time interval $[0,\tau]$ and be decoupled adiabatically for earlier and later times.
 For this we introduce a smooth coupling function $\alpha_{\tau}(t)$ 
that  takes the value  $1$ in the time interval $[0,\tau]$ and 
is zero beyond the  interval $[t_1,t_2]$ ($t_1<0 $ and $t_2>\tau$). The values for  $t_1<t<0$ and
$\tau<t<t_2$ are chosen in a way that  provides an  adiabatic switching.
  The entire Hamiltonian   reads then
\begin{equation}
 {H}(t)= {H}_{{\rm sys}}
 - \alpha_{\tau}(t){x}{A} + \frac{{q}^2}{2m} . 
\end{equation} 
 The Heisenberg equation of motion for the detector momentum
${q}$
\begin{equation} \label{eq:motion}
\dot{{q}}(t)=\alpha_{\tau}(t) {A}(t) 
\end{equation}
suggests, that the statistics of outcomes of  measurements of the detector's momentum after having it uncoupled from the system corresponds to the statistics of the time average  $\int_0^{\tau} dt \, A(t)$ that we are interested in.

The coupling term can be viewed as a disturbance of the system
measured by the detector.
To minimize this disturbance,
 we would clearly like to concentrate the detector wave function
 around $x=0$.  The uncertainty principle forbids us,
 however, to localize  it completely. Thereby we  would
  loose all information about the detector momentum,
 which we intend to measure.
  This is a fundamental limitation imposed by quantum mechanics,
  and we are going to explore
  its consequences step by step. To discern it form a classical back action of the detector we take the limit of a static detector with $m \to \infty$, such that  $\dot{{x}}=0$ and
  any classical  back action is ruled out.

\subsection{ Approach }
\label{sc:staticcalculation}

To predict the statistics of measurement outcomes in our detection model we need the reduced density matrix of the detector after the measurement, at $t>t_2$. If there were no system to measure
we could readily express it in the form of a path
integral in the (double) variable $x(t)$ over the exponential of the detector action.  This is still possible in the presence of a system coupled to the detector \cite{Fey63}. The information about the system to be measured can be compressed
into an extra factor in this path integral, the so-called influence functional.
 This makes  the separation between the detector and the
measured system explicit. To make contact with  \cite{Lev93},
 we write  the influence functional
as an operator expression that involves  system degrees of freedom 
only.
We denote the initial detector density matrix  (at $t<t_1$) by
$\rho^{in}(x^+,x^-)$ and the final one (at $t>t_2$, after having traced out the system's degrees of freedom) by
 $\rho^{f}(x^+,x^-)$. ${R}$
  denotes the initial 
 density matrix of the system.
  The entire initial density matrix $D$ is assumed to factorize, ${D}={R}{\rho}^{in}$.
     
 We start out by inserting complete sets of states into
      the expression for the time development of the density matrix  
\begin{eqnarray}
{\rho^{f}}(x^+,x^-) &=& \mathop{Tr}\limits_{\rm System} \, \langle x^+|\overrightarrow{T}e^{-i \int_{t_1}^{t_2}{dt\;
\bigl[  {H}_{\rm sys}-\alpha_{\tau}(t){x} {A}+\frac{{q}^2}{2m} \bigr]}}\;  {D} \nonumber \\
&&\;\;\;\;\;\;\; \overleftarrow{T}e^{{i}
\int_{t_1}^{t_2}{dt\;
 \bigl[{H}_{\rm sys}-\alpha_{\tau}(t){x} {A}+\frac{{q}^2}{2m}\bigr] }}|x^-\rangle.
\end{eqnarray}
 Here, $\overrightarrow{T}(\overleftarrow{T})$ denotes  (inverse)
 time ordering. As the complete sets of states we choose  product
 states of  any complete set of states of the system and alternatingly complete sets of eigenstates of the position or the momentum operator of the detector.   Those intermediate states allow us to replace the position and momentum
 operators in the time development exponentials by their eigenvalues.
 We can then do the integrals over the system states as well as the momentum
 integrals and arrive at the expression
\begin{eqnarray} \label{eq:staticresult}
 &&{\rho^{f}}(x^+,x^-) = \int{ \mathop{{\cal D}[x^+]}\limits_{x^+(t_2)=x^+} \mathop{{\cal D}[x^-]}\limits_{x^-(t_2)=x^-} \,\rho^{in}[x^+(t_1),x^-(t_1)]\; e^{-{i}   {\cal S}_{\rm Det}([x^+],[x^-]) }} \nonumber \\
 &&\mathop{Tr}\limits_{\rm System}  \,\overrightarrow{T}e^{-{i} \int_{t_1}^{t_2}{dt  \bigl[ {H}_{\rm sys}-\alpha_{\tau}(t)x^+(t) {A}\bigr] }}\,  {R} \, \overleftarrow{T}e^{{i} \int_{t_1}^{t_2}{dt \bigl[ {H}_{\rm sys}-\alpha_{\tau}(t)x^-(t) {A} \bigr] } }  
\end{eqnarray}
with the detector action
\begin{equation}
   {\cal S}_{\rm Det}([x^+],[x^-]) = - \int_{t_1}^{t_2}{dt\, \frac{m}{2} \bigl[ (\dot{x}^+)^2-(\dot{x}^-)^2\bigr]}. 
\end{equation}
We rewrite the expression as 
\begin{equation}
\rho^{f}(x^+,x^-) = \int{dx_1^+ dx_1^- \;K(x^+,x^-;x_1^+,x_1^-,\tau) \rho^{in}(x_1^+,x_1^-)}
\end{equation}
with the kernel 
\begin{eqnarray}
 &&K(x^+,x^-;x_1^+,x_1^-,\tau) = \int{ \mathop{{\cal D}[x^+]}\limits_{x^+(t_2)=x^+, x^+(t_1)=x_1^+} \;\;\mathop{{\cal D}[x^-]}\limits_{x^-(t_2)=x^-, x^-(t_1)=x_1^-}} \nonumber \\
\label{eq:kernel}
 &&{\cal Z}_{\rm Sys} ([\alpha_{\tau} x^+],[\alpha_{\tau} x^-])\, e^{-{i}  {\cal S}_{\rm Det}([x^+],[x^-])}
\end{eqnarray}
that contains  the  influence functional
\begin{equation}
 {\cal Z}_{\rm Sys} ([\chi^+],[\chi^-]) =  \mathop{Tr}\limits_{\rm System}
 \overrightarrow{T} e^{-{i} \int_{t_1}^{t_2}{dt \bigl[{H}_{\rm sys}-\chi^+(t) {A}\bigr]}}\;  {R} \; \overleftarrow{T}\; e^{{i} \int_{t_1}^{t_2}{dt \bigl[ {H}_{\rm sys}-\chi^-(t) {A}\bigr] }} .
\label{eq:time_dependent}
\end{equation}
Taking the limit of an infinite detector mass
now, we find that  ${\cal S}_{\rm Det}$ in Eq.\ (\ref{eq:kernel})
 suppresses all fluctuations in the path integral.
Therefore, the kernel $ K(x^+,x^-,x_1^+,x_1^-,\tau)$
becomes local in position space, 
\begin{equation}
 K(x^+,x^-,x_1^+,x_1^-,\tau)= \delta(x^+ - x_1^+)\; \delta(x^- - x_1^-) \;\; P(x^+,x^-,\tau) 
\end{equation}
with
\begin{equation}  \label{eq:local}
 P(x^+,x^-,\tau) =  \mathop{Tr}\limits_{\rm System}
  \overrightarrow{T} e^{-{i} \int_{t_1}^{t_2}{dt \bigl[ {H}_{\rm sys}-\alpha_{\tau}(t)x^+
  {A} \bigr] }}
   {R}  \overleftarrow{T} e^{\;{i} \int_{t_1}^{t_2}{dt \bigl[ {H}_{\rm sys}-\alpha_{\tau}(t)x^-{A} \bigr]}} .
\end{equation}

It is constructive to rewrite now the
density matrices in the  Wigner representation
\begin{equation}
 \rho(x,q) = \int{\frac{dz}{2 \pi} \; e^{-iqz}\; \rho(x+\frac{z}{2},x-\frac{z}{2})}
\end{equation}
and define
\begin{equation} \label{eq:P(x,q)}
 P(x,q,\tau)=  \int{ \frac{dz}{2 \pi} e^{-iqz}\;P(x+\frac{z}{2},x-\frac{z}{2},\tau)}.
\end{equation}
With these definitions we find a  compact  relation that summarizes the results of this
subsection:
\begin{equation} \label{eq:dens}
 \rho^f(x,q) = \int{dq_1 \; P(x,q-q_1,\tau)
 \; \rho^{in} (x,q_1)}.
\end{equation}

\subsection{Interpretation of the  FCS}
\label{Interpretation}
We adopt the relations (\ref{eq:local}),
(\ref{eq:P(x,q)}) and (\ref{eq:dens}) as the definition of
the FCS of the variable $ A$. Let us see why. First let us suppose
that we can  treat the detector classically. Then the density
matrix of the detector in the Wigner representation
can be interpreted as a classical probability distribution 
$\Pi(x,q)$ to be at a certain position $x$  with momentum $q$.
This allows for a classical interpretation of $P(x,q,\tau)$
as the probability to have measured  $q = \int_0^{\tau} A(t)$.
Indeed, one sees from (\ref{eq:dens}) that  
$\rho^f(x,q)$ is obtained from $\rho^{in}$ by shifts in $q$,
$P(x,q,\tau)$ being the probability distribution of those shifts.

In general, the  density matrix in the Wigner representation
cannot be interpreted as a probability to have a certain position
and momentum since it is not positive.
Concrete calculations given below illustrate that  $P(x,q,\tau)$
does not have to be positive either. Consequently, it cannot
be interpreted as a probability distribution.  Still it predicts the results
of measurements  according to  Eq.\ (\ref{eq:dens}). 

There is, however, an important case when the FCS can indeed
be interpreted as a probability
distribution. It is the case that  $P(x,q,\tau)$ does not depend on $x$,
$ P(x,q,\tau) \equiv P(q,\tau)$. Then, integrating   Eq.\  (\ref{eq:dens})  over $x$, we find
\begin{equation}
\Pi^f(q)=\int{dq' \; P(q-q',\tau)\; \Pi^{in}(q')}
\end{equation}
with $ \Pi(q) \equiv \int{dx \, \rho(x,q)}$. 
Therefore, the FCS is in this special case the kernel that relates the probability distributions of the detector momentum before and after the measurement, $\Pi^{in}(q)$  and $\Pi^{f}(q)$, to each other. Those distributions are positive and so is  $P(q,\tau)$.

 
When studying the  FCS of a stationary system and the measurement
time $\tau$ exceeds time scales associated with the system,
the operator expression in Eq.\ (\ref{eq:local}) can be seen
as a product of terms corresponding to time intervals.
Therefore  in this limit of $\tau \rightarrow \infty$
the dependence on the  measuring time
can be reconciled into
\begin{equation}
P(x^+,x^-,\tau)= e^{-{\cal E}(x^+,x^-) \tau}
\end{equation}
where the expression in the exponent is supposed to be large.
Then the  integral (\ref{eq:P(x,q)}) that defines the FCS can be done in
the saddle point approximation. Defining the time average $\bar{A}=q/\tau$,
that is,  $\bar{A}$ is the result of a measurement of $\int_0^{\tau} A(t) dt/\tau$,
the FCS can be recast into the form
\begin{equation}
P(x,\bar{A},\tau) = e^{-{\tilde{\cal E}}(x,\bar{A}) \tau},
\end{equation}
where ${\tilde{\cal E}}$ is defined as the (complex)
extremum with respect to (complex) $z$:
\begin{equation}
{\tilde{\cal E}}=\mathop{{\rm extr}}_{z} \{ {\cal E}(x+\frac{z}{2},x-\frac{z}{2}) + i \bar{A} z\}.
\end{equation}
The average value of $\bar{A}$ and its variance (noise) can be expressed
in terms of derivatives of ${\cal{E}}$:
\begin{eqnarray}
\langle\bar{A}\rangle& =&- \lim_{z \rightarrow 0} \frac{\partial {\cal E}
(x+z/2,x-z/2)}{i \partial{z}} \nonumber \\
  \tau \langle \langle \bar{A}^2 \rangle\rangle & =& \lim_{z
\rightarrow 0} \frac{\partial^2 {\cal E}
(x+z/2,x-z/2)}{\partial{z^2}}. 
\end{eqnarray}
More generally, the quantity $P(x^+,x^-,\tau)$ is  the generating
function of moments of $q$. It is interesting to note
that in general this function may generate a variety
of moments that differ in the time order of operators involved,
for instance,
\begin{eqnarray}
Q^N_M = (-1)^M i^{N} \lim_{x^\pm \rightarrow 0}
\frac{\partial^M} {\partial (x^-)^M}
\frac{\partial^{N-M}} {\partial (x^+)^{N-M}} P(x^+,x^-,\tau) \nonumber \\
=\int_0^\tau dt_1 ... d t_N\,
\langle\overleftarrow{T} \{ {A}(t_1)...{A}(t_M)\}
\overrightarrow{T} \{{A}(t_{M+1})...{A}(t_N)\}\rangle.
\end{eqnarray}
The moments of (the not necessarily positive) $P(0,q,\tau)$ are expressed through these moments
and binomial coefficients,
\begin{equation}
Q^{(N)} \equiv \int dq\, q^N P(0,q,\tau) = 2^{-N} \sum_M {N\choose M} Q^N_M.
\end{equation}
For $A=I$ these moments correspond to the moments generated by the $\chi(\lambda)$ of the previous section, that is contained in $P(x^+,x^-,\tau)$ as $\chi(\lambda)=P(\lambda/2,-\lambda/2,\tau)$ [compare Eqs.\ (\ref{eq:chic}) and (\ref{eq:chifromchic}) in the semiclassical approximation, where $H_{\lambda}= H_0 - \lambda I$, to Eq.\ (\ref{eq:local})].
 Interpreting $\chi(\lambda)$ as the characteristic function corresponding to a probability distribution is therefore equivalent to the classical interpretation of $P(x,q,\tau)$ discussed above. It is applicable to systems with an $x$- independent $P(x,q,\tau)$. 

\subsection{FCS of a system in the ground state}
 To acquire a better understanding of the general relations
 obtained we consider now an important  special case.
 We will assume that the system considered is in its
 ground state $|g\rangle$, so that its initial density
 matrix is ${R}=|g\rangle\langle g|$. In this case   the FCS is easily calculated.
  We have assumed that the coupling between the system
  and the detector is switched on  adiabatically.
  Then the time
  development operators in (\ref{eq:local}) during the 
  time interval $t_1<t<0$ adiabatically transfer
  the system  from $|g\rangle$ into the ground state $
  |g(x^{\pm})\rangle$ of the new Hamiltonian
  ${H}_{\rm sys}-x^{\pm}{A}$ .
  In the time interval $0 < t < \tau$
  the time evolution of the resulting state has then the
   form 
\begin{equation}
 e^ { - {i}\;t\;({H}_{\rm sys}-x^{\pm}\,{A})}\;
|g(x^{\pm})\rangle\;=\; e^{-{i}tE(x^{\pm})}\;|g(x^{\pm})\rangle.
\end{equation}
 Here, $E(x^{\pm})$ are the energies corresponding
 to $|g(x^{\pm})\rangle$.
 This gives the main contribution to the
 FCS if the measurement time is large and the phase acquired
 during the switching of the interaction
 can be neglected in comparison with this contribution,
\begin{equation} \label{eq:ground}
  P(x^+,x^-,\tau) =  e^{- {i} \tau [E(x^+)-E(x^-)]}.  
\end{equation}
We now  assume the function $E(x)$ to be analytic
and expand it in its Taylor series. We also re-scale $q$ as above, $\bar{A}=q/\tau$. We have then  for the FCS
\begin{equation}  \label{eq:Pground}
 P(x,\bar{A},\tau)=  \tau \int{dz\; e^{-iz \bar{A} \tau}\cdot  e^{-{i} \tau[ E'(x) z + E'''(x) z^3/24 + ...]}}.
\end{equation}
First we observe that $P(x,\bar{A},\tau)$ is a real function
in this case, since the exponent in (\ref{eq:Pground})
is anti-symmetric in $z$.
A first requirement for being able to interpret it
as a probability distribution is therefore fulfilled.
However, the same asymmetry assures that all {\it even}
cumulative moments of $\bar{A}$ are identically zero, whereas
the odd ones need not. On one hand, since the second
moment corresponds to the noise and the ground state cannot
provide any, this makes sense. On the other hand,
this would be impossible if $P(0,\bar{A},\tau)$
were a  positive probability distribution unless it had no dispersion at all.

Belzig and Nazarov \cite{Bel01} encountered this situation
analyzing the FCS of a  super-conducting junction.
In a certain limit the junction becomes a 
Josephson junction in its ground state. In this  limit
the interpretation of the FCS as a probability distribution
does not work any longer.
Fortunately enough,  relation
(\ref{eq:dens}) allows us to interpret
the results obtained.

In the limit $\tau \rightarrow \infty$ of Eq.\ (\ref{eq:Pground})  all
the terms involving higher derivatives of $E(x)$ are negligible and we  have
\begin{equation} \label{eq:Pinfinite}
 \lim_{T\to\infty} \,P(x,\bar{A},\tau)= \delta[\bar{A} + E'(x)].
\end{equation}
According to the Hellman-Feynman theorem $E'(x)=-\langle g(x)|{A}|g(x)\rangle$.
As one would expect, in this limit the measurement gives the expectation value
of the operator ${A}$ in a ground state of the system
that is somewhat altered by its interaction  with the detector
at  position $x$. Therefore the  resulting dispersion of $A$ will be determined
by the {\it initial} quantum mechanical spread of the detector wave function.
The error of the measurement stems from the interaction with
the detector rather than from  noise intrinsic in the  measured system.

\subsection{FCS of  electric current in a normal conductor} \label{sc:normal}
A complementary example is a normal conductor biased at finite
voltage. This is a stationary { \it non-equilibrium} system far from being
in its ground state. Here we do  not go to microscopic details
of the derivation. Our immediate aim is to make contact with the approaches of
Refs. \cite{Lev93,Naz99}. We keep the  original notations of the references wherever
it is possible.

The FCS of the current
in a phase-coherent conductor  is characterized by a set
of transmission coefficients $\Gamma_n$ [c.\ f.\  Eq.\ (37) of \cite{Lev93}]. At zero temperature and bias voltage $V$ it  reads
\begin{equation}
\label{eq:zsys}
P(x^+,x^-,\tau) = \exp \left\{  \frac{  e \tau}{2\pi}  | V |  S[i e(x^+-x^-){\rm sign} V ] \right\}
\end{equation}
with the function
\begin{equation}
 S(\xi)=\sum_{n}\ln\bigl[1+(e^{\xi}-1)\Gamma_{n}\bigr].
 \label{eq:sofxi}
\end{equation}
This expression depends on $x^+ - x^-$ only, as  a
direct consequence of gauge invariance. Indeed, in every time development operator of Eq.\ (\ref{eq:local}) (with $A=I$) the coupling term is the coupling to a vector potential
localized in a certain cross-section of the conductor. The gauge
transform that shifts the phase of the wave functions by $e x^{\pm}$
on the right side of that cross-section eliminates this coupling term.
This transformation was explicitly implemented in \cite{Naz99}.
Since there are {\it two} time development operators  in the expression, the coupling
terms cannot be eliminated simultaneously when  $x^{+} \neq x^{-}$.
However, the gauge transform with the phase shift $ e(x^{+} + x^{-})/2$
makes the coupling terms depending on $x^+ - x^-$ only.

Since $P(x^+,x^-,\tau)$ depends on $x^+ - x^-$ only,
the FCS $P(x,q,\tau)$ is independent of  $x$. As we have seen
in section \ref{Interpretation}, under this condition the  FCS can be interpreted 
as a probability distribution. As in section \ref{sc:elementary} we conclude that for the statistics of the current of  non-interacting electrons the characteristic function $\chi(\lambda) = P(\lambda/2,-\lambda/2,\tau)$ is associated with a positive probability distribution. 

Superconductivity breaks gauge invariance, thus making such an interpretation
impossible.

 \section{Electric Circuits as General Linear Detectors} \label{sc:linear}

The model used in the previous section may seem  rather abstract and
unrealistic. To make contact to the "real world", we notice that
\begin{itemize}
\item the time derivative of the detector 
momentum $q$ is related to the current 
through the mesoscopic system 
\item  the velocity $\dot x$ enters the Hamiltonian
in the same way as the voltage applied to the conductor. 
\end{itemize}
Next we adopt
the following definition of the "real world": the only quantities measured
are electric voltages and currents between nodes of an electric circuit.
The most adequate description of the  quantum mechanics of the system would thus
contain
these variables only. This description is hardly possible to achieve within
a Hamiltonian formalism, since the latter can contain neither dissipation nor
retardation. It is the Feynman-Vernon formalism \cite{Fey63}
that allows to formulate quantum mechanics in the form of an action that
contains only necessary variables. 
This action may be derived from the Hamiltonian formalism by tracing out 
irrelevant degrees of freedom.
But this is precisely what we have done in the previous sections!
The conclusion is that the above results can be used to arrive at an adequate
formulation of the  quantum dynamics of electric circuits that contain mesoscopic 
conductors. This formulation clarifies the notion of a detector. 

\begin{figure} \centering\epsfig{file=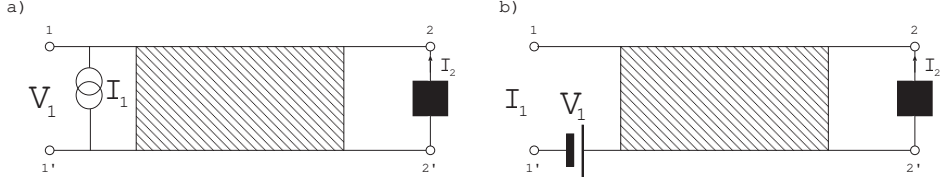,clip=,width=\linewidth}
\caption{
Electric circuit as a general linear detector.
The black box
symbolizes the electronic system to be measured.  It is embedded in
a linear environment (shaded
box). Either
 voltage (a) or current
(b) is  measured between
  the output contacts $1$ and $1'$ of the
circuit.} \label{fig1} \end{figure}

We formulate the  problem  as follows. Suppose we know
the FCS of a mesoscopic conductor. When measured, it is embedded 
in an electric circuit.  Generally speaking, one does not measure 
the voltage or current directly at the mesoscopic conductor  
but rather
somewhere else in the circuit. 
We cast  this in Fig.\
\ref{fig1} in circuit theory terms. Either a voltage or a current
measurement is performed at the  output contacts $1$ and $1'$.
The mesoscopic
conductor (black box) is connected to the input contacts $2$ and $2'$.
The shaded box presents the electromagnetic environment of the
mesoscopic conductor. It is supposed to be a piece of
a linear circuit and thus can be characterized separately from the 
"black box" by three (frequency-dependent) response
functions. 
The question is what is the FCS of such a measurement.

We answer it by extending
our simple detection model of the previous section to a set of dynamical detectors. We model 
the "environment", including the   detectors, by a set of additional    degrees of freedom. This way we study classical back action effects of the detectors.
These effects will be more important than the subtle "quantum"
influence of the detector on the measured system discussed in the previous section. 
The detectors, representing the environment, shall
obey linear equations of motion. This restriction to linear circuits is
 compatible with  most  experimental situations. Detector non-linearities
are avoided in many experiments because of undesired
effects like the mixing of different frequencies. 

\subsection{The model of a linear electromagnetic environment}
\label{sc:currentmodel} 
Linear systems can be modeled by  a set of non-interacting
bosons \cite{CL}. The Hamiltonian  reads
\begin{equation}
H_{\rm linear} = \sum_m \Omega_m a_m ^{\dagger} a_m - \sum_i h_i(t) H_i 
\end{equation}
where the operators $H_i$ are the physical quantities of interest
and  conjugated to generalized forces $h_i$.
These operators are linear combinations of $a_m ^{\dagger}, a_m$
\begin{equation}
H_i = \sum_m c^{(i)}_m a_m + h.c. .
\end{equation} 
A proper choice of $\Omega_m, c^{(i)}_m$ models arbitrary response
functions and enables one to formulate the quantum dynamics of the circuit in terms
of quantities of interest only \cite{CL}.
We are dealing with two pairs of conjugated variables, $(\chi_1,I_1)$ and $(\chi_2,I_2)$, where  $\chi_j(t) =\int^{t} d\tau V_j(\tau)$.
One variable of each pair  can be chosen as generalized force and the remaining
one as  operator. Although different choices lead to different
Hamiltonians, the concrete choice is just a matter of convenience.
Our primary choice is to treat $I_2$ and $\chi_1$ 
as generalized forces.
The response functions relate $I_1$ and $V_2$ to those two variables:   
\begin{eqnarray}
I_1(\omega)&=&\tilde Y(\omega) V_1(\omega) + K(\omega) I_2(\omega) \nonumber \\
V_2(\omega)&=&\tilde Z(\omega) I_2(\omega) - K(\omega) V_1(\omega).
\end{eqnarray}
More symmetric choices express either currents in terms of voltages
via admittances $Y_1,Y_2,Y_{12}$,
\begin{eqnarray}
I_1(\omega)&=&Y_1(\omega) V_1(\omega) + Y_{12}(\omega) V_2(\omega) \nonumber \\
I_2(\omega)&=&Y_2(\omega) V_2(\omega) + Y_{12}(\omega) V_1(\omega),
\end{eqnarray}
or voltages in terms of currents via impedances $Z_1,Z_2,Z_{12}$,
\begin{eqnarray}
V_1(\omega)&=&Z_1(\omega) I_1(\omega) + Z_{12}(\omega) I_2(\omega) \nonumber \\
V_2(\omega)&=&Z_2(\omega) I_2(\omega) + Z_{12}(\omega) I_1(\omega). \label{eq:impedances}
\end{eqnarray}
There are obvious relations between these response coefficients:
$\tilde Z =1/Y_2, \tilde Y = (Y_1 Y_2 -Y_{12}^2)/Y_2, K=Y_{12}/Y_2$, 
$(Z_1,Z_2,Z_{12}) = (Y_2, Y_1, -Y_{12})/(Y_1 Y_2 - Y_{12}^2) $. 

Here we assume a passive circuit in thermal equilibrium.
The response functions then satisfy Onsager symmetry relations
and the fluctuation-dissipation theorem relates the response functions
to  fluctuations of the corresponding variables.

\subsection{General relation} \label{sc:approach}

We derive now the general relation
that  determines the full counting statistics of the output $I_1(t)$
of
the linear detector described above { \it provided} it is coupled
to the "black box" to be measured. We will
follow the lines of section \ref{sc:staticcalculation}. However, now
 we  assume the detector
to be in a state of thermal equilibrium at temperature $T$.
In addition, instead of
evaluating the final density matrix of the detector $\rho^f$,
we analyze  the partition functional
${\cal Z}_I$ that generates moments of the  read-off environment variable
$I_1$. We define  this functional  as
\begin{equation} \label{eq:Zcurr} {\cal Z}_I [ \vec{\chi}_1]=
\left\langle \overleftarrow{\rm T} e^{i \int{dt\; \left[{H}_I +
\chi_1^-(t) {I}_1\right]} } \overrightarrow{\rm T} e^{-i \int{dt\;
\left[{H}_I + \chi_1^+(t) {I}_1\right]} } \right\rangle.
\end{equation}
It generates moments of outcomes of measurements of $I_1(t)$ at
any instant of time.

It is advantageous to write the functional in
its dependence on the combinations  
$V^{cl}_1=(\partial/\partial t)(\chi^+_1 +\chi^-_1)/2$ and
$\chi^q_1=(\chi^+_1-\chi^-_1)/2$, that we collect into a "vector"
$\vec{\chi}_1= (V^{cl}_1 ,
\chi^q_1)$. The "classical" field $V^{cl}_1$ accounts for a 
(time-dependent) voltage source between the contacts $1$ and $1'$.
Moments of $I_1(t)$ are generated by differentiation of ${\cal Z}_I$
with respect to the anti-symmetric, sometimes
called "quantum", field:
\begin{eqnarray}
\label{eq:funcdiff}
\left\langle [I_1(t_1) ... I_1(t_m)] \right\rangle =
\frac{\delta}{ -2i \delta \chi_1^{q}(t_1)} ...
\frac{\delta}{ -2i \delta \chi_1^{q}(t_m)} { {\cal
Z}}_I[\vec{\chi}_1]\Big|_{{\chi^q}_1=0}.
\end{eqnarray}
In general, these moments  depend on $V^{cl}_1$.

${H}_I$ is the Hamiltonian of the
circuit  for a current
measurement. It reads
\begin{equation}
H_I = H_{\rm linear} + H_{\rm Sys} - \chi_2 I.
\end{equation}
$ H_{\rm linear}$ governs the bosonic detector degrees of freedom
and  $H_{\rm Sys}$ acts on the  electronic degrees of freedom
of the "black box". The third term  couples the  electric
current  to the detector degree of freedom
$\chi_2(t)= \int^t dt'\, V_2(t')$. The latter is thus the analogue of
$x$ in  section
\ref{sc:static}.

Both $\chi_2$ and $I_1$ are linear combinations
of boson creation/annihilation operators,
\begin{equation}
\chi_2 = \sum_m c^{(\chi)}_m a_m+ c_m^{(\chi) *} a_m
^{\dagger}, \ \ 
I_1  = \sum_m c^{(I)}_m a_m+ c_m^{(I) *} a_m
^{\dagger}. \end{equation}
We rewrite now ${\cal Z}_I$ as a path integral in detector
variables, like we have done to derive  Eq.\
(\ref{eq:staticresult}). The integration variables now
are $a^{(\pm)}_m(t)$, $\pm$ corresponding to the  two parts of the
Keldysh contour. Operators $\chi_2, I_1$ are replaced by
\begin{equation}
\chi_2^{(\pm)} = \sum_m c^{(\chi)}_m a^{(\pm)}_m+ h.c.
\ \ 
I_1^{(\pm)}  = \sum_m c^{(I)}_m a^{(\pm)}_m+ h.c., \end{equation}
the sign depending on the part of the contour they reside on.
To proceed, we introduce two extra vector variables into the path
integral: $\vec{\chi}_2 = (V^{cl}_2,\chi^q_2)$ and
$\vec{q}_2=(I^{cl}_2,q^{q}_2)$,
by inserting the identity 
\begin{eqnarray}
1& \simeq & \int  {\cal D}[V^{cl}_2]
{\cal D}[\chi_2^q]  \prod_t \delta (2V_2^{cl} -\dot \chi^+_2-\dot \chi^-_2)
\delta (2\chi_2^{q} - \chi^+_2+ \chi^-_2) \nonumber \\
&\simeq&  \int {\cal D}[V^{cl}_2]
{\cal D}[\chi_2^q] {\cal D}[I^{cl}_2]
{\cal D}[q_2^q] \nonumber \\
&& \exp \left\{i \int dt\, [I^{cl}_2(2\chi_2^{q} - \chi^+_2+ \chi^-_2)
-q_2^q(2V_2^{cl} -\dot \chi^+_2-\dot \chi^-_2)]\right\}.  
\end{eqnarray}
 This allows us to split the integrand into two factors. One factor
is a trace over "black box" variables. It depends on $\vec{\chi}_2(t)$
only and is in fact the influence functional ${\cal Z}_{\rm Sys}$
introduced in  section \ref{sc:staticcalculation},
Eq.\ (\ref{eq:time_dependent}) (with $A=I$). It generalizes our definition of the  FCS of the "black box" biased by the  voltage $V^{cl}(t)$ to time dependent arguments.
The other factor is a quadratic form in the  $a_m(t)$. Its Gaussian character  enables us
to perform the  integrations over  the  $a_m$, yielding
\begin{eqnarray}
\label{eq:result}
{\cal Z}_I[\vec{\chi}_1] &=&
\int {\cal D}[\vec{\chi}_2]{\cal D}[\vec{q}_2] e^{-i\{
{\cal S}_{\rm env}([\vec{\chi}_1],[\vec{q}_2] )
+{\cal S}_{\rm coup}([\vec{q_2}],[\vec{\chi_2}])\}}
{\cal Z}_{\rm Sys}[\vec{\chi}_2], \\
{\cal S}_{\rm env} &=& \int \frac{d\omega}{2\pi} \left(
 \vec{\chi_1}\check{\tilde {Y}}\vec{\chi_1}
+ 2 \vec{\chi_1}\check{K} \vec{q}_2
+\vec{q}_2  \check{\tilde{Z}} \vec{q}_2 \right)  \\ 
{\cal S}_{\rm coup} &=& 2 \int dt\left[q_2^q(t)V_2^{cl}(t)-I^{cl}_2(t)\chi_2^{q}(t) \right].
\end{eqnarray}
The environmental part of the action is expressed in terms
of response functions of our primary choice.
The $2\times 2$ matrices $\check{\tilde Y},\check K$ in the time domain
are integral kernels depending on the time difference only.
In frequency representation they read
\begin{eqnarray}
\check{\tilde Y}(\omega) & = &\left( \begin{array}{cc} 0 &
\tilde{Y}^*(\omega) \\
\tilde{Y}(\omega) & -2i\omega \coth(\frac{\omega}{2T}) {\rm Re} \, \tilde{Y}(\omega)
\end{array} \right), \label{Ymatrix}\\
\check{K}(\omega) & = &\left( \begin{array}{cc} 0 &
{-K}^*(\omega) \label{eq:groundmatrix}\\
K(\omega) & 2\omega \coth(\frac{\omega}{2T}) {\rm Im} \, K(\omega)
\end{array} \right).
\end{eqnarray}
The matrix ${\check {\tilde Z}}$ is of the same form as ${\check {\tilde Y}}$.
We can further simplify this relation by integrating over $\vec{q_2}$.
The result acquires the more symmetric form
\begin{eqnarray}
\label{eq:generalresult}
{\cal Z}_I[\vec{\chi}_1] &=&
\int {\cal D}[\vec{\chi}_2] e^{-i
{\cal S}_{\rm env}([\vec{\chi}_1],[\vec{\chi}_2]) }
{\cal Z}_{\rm Sys}[\vec{\chi}_2], \\
{\cal S}_{\rm env} &=& \int  \frac{d\omega}{2\pi} \left[
 \vec{\chi_1}\check{Y_1}\vec{\chi_1}
+ 2 \vec{\chi_1}\check{Y_{12}} \vec{\chi}_2
+\vec{\chi}_2  \check{Y_2}\vec{\chi}_2 \right],  \label{eq:linearaction}
\end{eqnarray}
where the matrices $\hat Y_1,\hat Y_2, \hat Y_{12}$ are defined as in Eq.
(\ref{Ymatrix}).
This is the desired general relation: it expresses the  FCS of electric currents through a  mesoscopic
conductor that is embedded in a linear electric circuit. It is important
that all information about the mesoscopic conductor  enters through  its FCS at voltage bias, ${\cal Z}_{\rm Sys}$.

Let us discuss the relation in some detail. First, let us replace
the general four-pole electric circuit by a single two-pole
resistor $Z(\omega)$ (see Fig. 2).
We do this by choosing $Y_1=Y_2=-Y_{12}=Y=1/Z$. The resistor and the "black
box" enter the expression then on equal footing,
\begin{eqnarray}
\label{eq:seriesresult}
{\cal Z}_I[\vec{\chi}_1] &=&
\int {\cal D}[\vec{\chi}_2]
{\cal Z}_{\rm Resistor}[\vec{\chi}_2-\vec{\chi}_1]
{\cal Z}_{\rm Sys}[\vec{\chi}_2], \\
{\cal Z}_{\rm Resistor}[\vec{\chi}] &=& \exp\left\{ -i\int  \frac{d\omega}{2\pi} 
 \vec{\chi}\check{Y}\vec{\chi}
  \right\}  ,
\end{eqnarray}
where $\check{Y}$ is defined as in Eq.\ ({\ref{Ymatrix}).
 ${\cal Z}_{\rm Resistor}$ defines the Gaussian FCS of the linear voltage-biased two-pole
resistor. Eq.\ (\ref{eq:seriesresult})
expresses  a simple concatenation rule for the  FCS of compound circuits:
the total FCS is a convolution of the FCS's of the  individual elements.
The resulting FCS is presented as an integral over the  fluctuating phase
$\vec{\chi}_2$
defined in one node of the circuit. This is in agreement with
known results about Keldysh actions of electric circuits with
Josephson junctions \cite{Schoen}. Moreover, the rule can be easily
generalized to more complicated circuits, for example, circuits with
two mesoscopic conductors in series.

\begin{figure}
\centering\epsfig{file=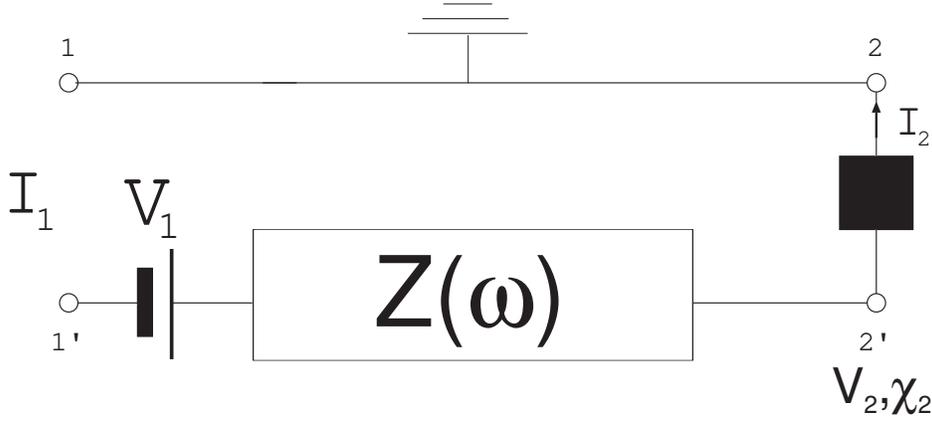,clip=,width= \linewidth}
\caption{
A simple example of a linear environment: a resistor $Z(\omega)$
in series with the "black box". The circuit is biased with a  voltage
source $V$. The  Quantum dynamics of the system is formulated
in terms of fluctuating fields $V_2,\chi_2$.
}
\label{fig2}
\end{figure}

It is interesting to  note that the integration over the fluctuating phase
in fact implements a constraint, namely  current conservation, on the quantum motion of the compound system.
 The current through the
"black box" equals
the input current of the detector:
\begin{eqnarray}
&&\delta(I_{\rm Sys} - I_{1}) \rightarrow
\prod_t \delta(I_{\rm Sys}(t)-I_1^+(t))
\prod_t \delta(I_{\rm Sys}(t)-I_1^-(t)) \simeq \nonumber \\
&&\int {\cal D}\vec{\chi}_2 \exp\left\{ i \int dt
[\chi_2^{+}(t)(I_{\rm Sys}(t)-I_1^+(t)) -
\chi_2^{-}(t)(I_{\rm Sys}(t)-I_1^-(t))] \right\}, \nonumber
\end{eqnarray}
the last equality holding for operators. Since the two currents are represented by operators of a different nature (fermionic  for $I_{\rm Sys}$ and bosonic for $I_1$), the
 constraint would be difficult to handle within a Hamiltonian formulation.
This illustrates the usefulness of the  Keldysh action approach in this context.

So far  we have disregarded the constant
factors that arise on various stages of the derivation
from the variable changes in the path integrals and/or
subsequent Gaussian integrations. 
This is not a matter of our carelessness but rather indicates
a  problem with the normalization in the approach in use. The normalization of  the final relation (\ref{eq:generalresult}) is  known
from the fact that $Z_I(\vec{\chi}) = 1$ at $\chi^q=0$
per definition of the  generating functional. In the path integral representation the correct normalization is assured by the causality structure of the action and integrals over the whole range of frequencies contribute   \cite{Alt00}. One can  not reproduce the correct factors within the low-frequency approximations and the saddle-point  treatment of the path integrals employed in this article. 
The problem is not related to the embedding of the mesoscopic
conductor and 
persists also to compound linear circuits.
One may  draw analogies with known, general and long-standing problems
related to the quantum mechanics of constrained motion \cite{constraines}.
To our knowledge it is always possible
to  correct for  wrong factors that arise in calculations
 by "normalizing" the results to $Z_I(\vec{\chi}) = 1$ at
$\chi^q=0$. For all general relations given in this article we assume that the correct factor is included
into the definition of the metrics in  function space, that is, into the 
definition of ${\cal D} [\vec{\chi_2}]$.

\subsection{Voltage measurement and Pseudo-Probabilities}
\label{sc:pseudo}

The evaluation of the FCS of electric current through a mesoscopic conductor with  voltage bias is a rather straightforward task.
One proceeds within the  
Hamiltonian formulation in terms
of electronic operators subjected to an external voltage.
The conjugated problem is the FCS of voltage fluctuations
under the condition of current bias. The absence of an obvious Hamiltonian
formulation for current bias  makes this problem less straightforward. 

Albeit the problem can be solved in a general way within the
approach of the previous subsection. Let us consider
the generating functional of  voltage fluctuations between 1 and 1' (Fig.
1 b) defined by 
\begin{equation} \label{eq:Zvolt}
 {\cal Z}_V [ \vec{q}_1]= \left\langle  
 \overleftarrow{\rm T} e^{i \int{dt\; \left[{H}_V + q_1^-(t) {V}_1\right]} }  \overrightarrow{\rm T} e^{-i \int{dt\; \left[{H}_V +  q_1^+(t) {V}_1\right]} } \right\rangle.
\end{equation}
Similarly to Eq. (\ref{eq:funcdiff}),  derivatives of this functional
with respect to $q^{q}(t)$ give the moments of voltage fluctuations.
We now repeat the derivation of
 the previous subsection for this
 functional. The answer reads:
\begin{eqnarray}
\label{eq:voltageresult}
{\cal Z}_V[\vec{q}_1] &=&
\int {\cal D}[\vec{q}_2] {\cal D}[\vec{\chi}_2] e^{-i\{
{\cal S}_{\rm env}([\vec{q}_1],[\vec{q}_2] )
+{\cal S}_{\rm coup}([\vec{q_2}],[\vec{\chi_2}])\}}
{\cal Z}_{\rm Sys}[\vec{\chi}_2], \\
{\cal S}_{\rm env} &=& \int  \frac{d\omega}{2\pi} \left[
 \vec{q_1}\check{Z_1}\vec{q_1}
+ 2 \vec{q_1}\check{Z}_{12} \vec{q}_2
+\vec{q}_2  \check{Z_2} \vec{q}_2 \right],  \\ 
{\cal S}_{\rm coup} &=& 2\int dt\left[q_2^q(t)V_2^{cl}(t)-I^{cl}_2(t)\chi_2^{q}(t) \right].
\end{eqnarray}
The matrices $\check{Z}_{1,2,12}(\omega)$ are defined analogously  to
(\ref{Ymatrix}) with the impedances  (\ref{eq:impedances}).

We now choose  the linear detector in such a way that the
voltage fluctuations between  1 and 1' are the same as between
2 and 2' and the conditions of current bias are
met. Elementary circuit theory tells us that this is achieved
in the limit $ Z_{12} \rightarrow -Z_1=-Z_2 \rightarrow -\infty$.
In the expression (\ref{eq:voltageresult}) this corresponds to
setting $\vec{q}_1 = \vec{q}_2$ and ${\cal S}_{\rm env} =0$.

This brings us to a remarkable conclusion:
for any conductor, the generating functionals for voltage noise
at current bias $I^{cl}(t)$ and for current noise at voltage bias
$V^{cl}(t)$ are
related by a functional Fourier transform,
\begin{equation}
\label{eq:voltcurr} {\cal Z}_V[\vec{q}] = \int{ 
{\cal D} [\vec{\chi}] \, e^{-2i \int{dt (q^{q} V^{cl} -
I^{cl} \chi^q)}} {\cal Z}_I[\vec{\chi}]}. \end{equation}
Therefore,
the functionals ${\cal Z}_V$ and ${\cal Z}_I$ are in fact just
different
forms of the same object. There is some uncertainty in this
definition stemming from the problem mentioned in the previous section.
 ${\cal Z}_V$ has to be "normalized" such that  ${\cal Z}_V =1$
at $q^{q}(t)=0$. We assume the normalization factor to be included
into the definition of ${\cal D}$.

This simple relation between ${\cal Z}_V$ and ${\cal Z}_I$ suggests to define the  functional 
\begin{eqnarray}
\label{eq:pseudo1}
\tilde{P}([V],[I]) &=& \int{ {\cal
D}[q^q] \,e^{2i\int{dt \, q^{q} V}} \;
{\cal
Z}_V \left[{I \choose q^q}\right]} \nonumber \\ &=& \int{ {\cal
D}[\chi^q] \,e^{2i\int{dt \, \chi^{q} I}}\;
{\cal Z}_I\left[{V \choose \chi^q}\right]} \end{eqnarray}
that depends on "classical" variables only. 
We show with
the help of Eq.\ (\ref{eq:funcdiff}) that $\tilde{P}$
 has the  following properties:
\begin{equation}
\label{eq:voltint} \left\langle V(t_1) ... V(t_m)
\right\rangle\Big|_{I(t)} =\frac{ \int{{\cal D}[V]\, V(t_1) ... V(t_m)}
\tilde{P}([V],[I])}
{\int{{\cal D}[V]\,
\tilde{P}([V],[I])}},
\end{equation}
and 
 \begin{equation}
\label{eq:currint} \left\langle I(t_1)...I(t_m) \right\rangle\Big|_{V(t)} =
\frac{\int{{\cal
D}[I]\,  I(t_1)...I(t_m) \tilde{P}([V],[I])}}{\int{{\cal D}[I]\, \tilde{P}([V],[I])}}
.
\end{equation}
Eqs.\ (\ref{eq:voltint}) and  (\ref{eq:currint}) resemble the properties of a probability density and yet $\tilde{P}$  is not a probability.
First, it does not have to be positive,
just as the $P(x,q,\tau)$  discussed in section \ref{sc:static} needs not.
Second,  $\tilde{P}$ is dimensionless in contrast to a true
probability density of either voltage or current fluctuations.
For these
reasons we refer to $\tilde{P}$ as a 'pseudo-probability'.

$\tilde{P}([V],[I])$ is a characteristics of a two-pole
conductor and takes a simple form for a linear circuit.
In this case $\tilde{P}$ is  positive
and depends only on the impedance of the conductor
$Z(\omega)$ and its temperature $T$, 
\begin{equation} \label{eq:Penv}
 \tilde{P}_{Z}([V],[I]) =
\exp\left\{ - \int{\frac{d\omega}{4\pi} \,
\frac{|V(\omega)-Z(\omega)I(\omega)|^2 }
{\omega {\rm Re} Z}
\tanh \frac{\omega}{2T}} \right\}.
\end{equation}
From Eq.\ (\ref{eq:voltint}) one derives the voltage correlator
\begin{equation} \label{eq:noiseex}
\left\langle V(t)^2 \right\rangle  =
\int{\frac{d \omega}{2 \pi}\; \omega \,{\rm Re} \,Z(\omega)\; {\rm coth} \frac{ \omega}{2 T}},
\end{equation}
that conforms to the fluctuation-dissipation theorem.

The general relation (\ref{eq:generalresult})
can be rewritten in terms of  pseudo-probabilities.
In this case, it will express the pseudo-probability
of current/voltage fluctuations between 1 and 1' $\tilde{P}_1([V],[I])$
in terms
of the $\tilde{P}_{\rm Sys}$
 of the "black box" and a four-pole
pseudo-probability $\tilde{P}_{12}$ that depends on two currents
and voltages, characterizing the linear part of the
circuit:
\begin{equation}
\tilde{P}_1([V_1],[I_1])=
\int{{\cal D}[V_2] {\cal D}[I_2]
\tilde{P}_{12}([V_1],[I_1];[V_2],[I_2])
\tilde{P}_{\rm Sys}([V_2],[I_2])}.
\end{equation}

We cast now also the result  Eq. (\ref{eq:seriesresult}) for
the "black box" in series with a linear resistor $Z(\omega)$  in terms of pseudo-probabilities. The answer
is expressed in terms of the pseudo-probability of the resistor
defined by Eq. (\ref{eq:Penv}):
\begin{equation}
\tilde{P}_1([V],[I])=
\int{{\cal D}[V_2] 
\tilde{P}_{Z}([V-V_2],[I])
\tilde{P}_{\rm Sys}([V_2],[I])}.
\end{equation}
The above relations are transparent if one interprets them
in terms of classical probabilities. They  show that the
probability of a certain current/voltage fluctuation is composed
of a probability of  fluctuations in the "black box" and of a probability of fluctuations  in the  linear
detector, voltage and current satisfying  circuit theory rules. 
Yet the relations are quantum mechanical ones and are written
for pseudo-probabilities. They contain
all necessary information about quantum properties of the system
under consideration.

To summarize,  fluctuations in a mesoscopic system that is
embedded in a  linear environment can be characterized by three statements. First,  voltages and currents
are related by circuit theory rules. For instance,
for the circuit of Fig. 2 the
current fluctuation $\delta I$ in the "black box" results in the
current fluctuation $ \delta I/(1 + ZG)$ in the whole circuit,
$G$ being the linear conductance of the "black box".
This factor is rather
trivial from a classical point of view. However, in the context of quantum mechanical detection, it constitutes the main source
of  detector back-action. This "classical" back-action
could be disregarded for the static detector considered
in section \ref{sc:static}.
Second, the linear part of the circuit produces its own Gaussian quantum
and thermal fluctuations. Also these fluctuations contribute to  the FCS of the compound system.
Third, the fluctuations in the linear environment affect the
fluctuations produced by the "black box".  This effect is the  least trivial
and we will explore its consequences  in section \ref{sc:QPC}.

\section{Low-frequency limit} \label{sc:low}

We have shown in the previous section that the FCS of a mesoscopic
conductor in a general linear environment can be expressed by the relation
(\ref{eq:generalresult}). This relation involves path integration  and defines  a quantum field theory with the field $\vec{\chi}_2(t)$.
This field theory is 
non-linear, the non-linearities coming from the mesoscopic
conductor. The latter is sometimes underestimated, since the  $I-V$ curve
of the  conductor may be perfectly linear. However, even in this
case the noise and higher order correlators in general do depend on the voltage, which
provides the non-linearity.

 Such non-linearities make
the general problem already hardly tractable. To complicate the situation further,
the most important part of the action, $\ln {\cal Z}_{\rm Sys}$,
is only known for {\em stationary} $\vec{\chi}$. In this case it is given
by the relation (\ref{eq:zsys}). Microscopic Keldysh Green function techniques
can be used to evaluate this functional at non-stationary $\vec{\chi}$. The answer has, however, a complicated functional dependence on the argument $\vec{\chi}$. 

Fortunately enough, there is a simple low-frequency  limit where the quantum field
theory becomes tractable. This limit applies to many  experimental situations. In this 
limit we consider only quasi-stationary realizations of the fields.
The action for these realizations is {\it local} in time, which allows for
easy path integration.

Before turning to the  concrete formulation, let us discuss the time scales
that occur in our problem and that  may determine the
non-locality of the action.
There may be a time scale, the $RC$-time, characterizing the  frequency dependence of
the detector response functions $Y_{1,2,12}$. This scale depends
on the system layout and can be easily changed. From
the experimental point of view it is convenient
to choose the response functions to be constant in a wide frequency
interval. This allows us to disregard this time scale.
Another time scale, $\tau_{Q}$, has  quantum mechanical origin.
The environment action (\ref{eq:linearaction}) becomes non-local
at frequencies $\omega \simeq k_B T$, where the crossover from classical
to quantum noise  occurs.
For the mesoscopic conductor a similar scale may arise from
the energy  $eV$ related to single electron transfer through the
conductor. Therefore $\tau_Q \simeq \hbar/{\rm max}\{eV,k_BT\}$.
A third time scale is the typical time interval $\tau_I$ between
 electron transfers through the mesoscopic conductor, $e/I$.
Comparing the latter two scales, we find that $t_I \simeq (e^2/h G)
\tau_Q$, $G$ being the conductance of the "black box".
Generally, we expect the low-frequency limit to be valid
at time scales exceeding both $\tau_I$ and $\tau_Q$.
These time scales are typically short in comparison
with the  time resolution of the  measurement electronics.
Consequently most electric
measurements are performed in the low-frequency limit.

In this limit, the general relation
(\ref{eq:generalresult}) takes the following form:
\begin{eqnarray}
Z_I([V_1],[\chi_1]) &=& \int {\cal D}[V_2] {\cal D}[\chi_2]
\exp\Big\{-\int dt\, [{\cal E}_{\rm
env}(V_1(t),\chi_1(t);V_2(t),\chi_2(t))
\nonumber \\ 
&&\;\;\;\;\;\;\;\;\;\;\;\;\;\;\;\;\;\;\;\;\;\;\;\;\;\;\;\;\;\;\;\;\;\;\;\;\;\;\;\;\;\;\;\;\;+ {\cal E}_{\rm Sys} (V_2(t),\chi_2(t))]\Big\}; \;\;\;\;\;
\end{eqnarray}
where 
\begin{eqnarray}
{\cal E}_{\rm env}(V_1,\chi_1;V_2,\chi_2) &=& -2i[(V_1 Y_1 +V_2 Y_{12})
\chi_1 +(V_2 Y_2 +V_1 Y_{12})
\chi_2] \nonumber \\
&&  +4T (Y_{1}\chi_1^2 +Y_2  \chi_2^2  +  2Y_{12} \chi_1\chi_2), \;\;\;\;
\nonumber \\
{\cal E}_{\rm Sys} &=& -{\rm ln} \; Z_{\rm Sys}(V,\chi)/\tau. \nonumber
\end{eqnarray}
In the last equation, $Z_{\rm Sys}$ is evaluated for stationary
$V,\chi$. Thus its logarithm  is proportional to  the measurement time $\tau$
(see Eq. (\ref{eq:zsys})). For the resistor in series (Fig. 2)
${\cal E}_{\rm env}$ reduces to:
\begin{equation}
{\cal E}_{\rm env}= {\cal E}_Z(V_1-V_2,\chi_1-\chi_2);
\; {\cal E}_Z(V,\chi) = (-2i V +4T \chi)\chi/Z.
\end{equation}
As expected in the low-frequency limit,
the noise of the detector is just Johnson-Nyquist white noise.

A simple estimation shows that the action is not only local in
time. It is also big for the time intervals in question. This allows
us to implement a semiclassical approximation, that is, to evaluate
the path integrals in saddle-point approximation. Indeed,
the typical values of $\chi$, manifesting the discreteness of charge,
are of the order of $1/e$. Typical values of the action are then estimated
as $ S \simeq I \chi \tau \simeq I/e \, \tau \simeq (\tau/\tau_I)$.
By definition of the  low-frequency limit, $\tau / \tau_I \gg 1$.

We now come  back to the central problem  addressed in this article:
the distribution of  currents averaged over the time interval $\tau$. Taking the path integrals
in  saddle-point approximation, we arrive at a simple expression
for the  FCS of the "black box" embedded in an environment circuit:
\begin{eqnarray}
Z_I(V_1,\chi_1)& =& \exp[-{\cal E}_I(V_1,\chi_1)\tau], \nonumber \\
{\cal E}_I(V_1,\chi_1)& =& \mathop{{\rm extr}}_{V_2,\chi_2} \left\{
{\cal E}_{\rm env}(V_1,\chi_1;V_2,\chi_2) + {\cal E}_{\rm
Sys}(V_2,\chi_2)\right\}.
\label{eq:simple}
\end{eqnarray}
 
The corresponding FCS of voltage fluctuations at given current $I$
can be obtained directly
from Eq. (\ref{eq:simple}) by using Eq. (\ref{eq:voltcurr}),
\begin{eqnarray}
Z_V(I,q) &=& \exp[-{\cal E}_V(I,q)\tau], \nonumber \\
{\cal E}_V(I,q) &= & \mathop{{\rm extr}}_{V,\chi} \left\{
{\cal E}_{I}(V,\chi) -2i(I\chi - qV)\right\}.
\label{eq:simplevolt}
\end{eqnarray}

When integrating Eq. (\ref{eq:voltcurr}) we take care of the  normalization
to assure that $Z_V(I,0)=1$. A similar integration
determines the pseudo-probability:
\begin{eqnarray}
\tilde{P}(I,V) &=& \exp(-{\cal E}(V,I)\tau), \nonumber \\
{\cal E}(I,V)& =& \mathop{{\rm extr}}_{\chi} \left\{
{\cal E}_I(V,\chi) - 2i I \chi \right\}.
\label{eq:simplepseudo}
\end{eqnarray}
Finally, we give here expressions for real probabilities
to measure the  voltage $V$ at a given current $I$, $P_V(I)$, and
to measure the  current $I$ at a given voltage $V$, $P_I(V)$.
By virtue of the  relations (\ref{eq:voltint}) and (\ref{eq:currint})
we obtain
\begin{equation}
\left\{\begin{array}{c}  P_V(I) \cr P_I(V) \end{array} \right\} =
\sqrt{\frac{\pi}{2 \tau}}
\left\{\begin{array}{c} \sqrt{\frac{\partial I^2} {\partial^2 {\cal E}}}
\cr \sqrt{\frac{\partial V^2} {\partial^2 {\cal E}} }\end{array}
\right\} \tilde{P}(I,V).
\label{eq:probabilities}
\end{equation}
Remarkably, these probabilities are in fact {\it the same} with
exponential accuracy. They differ in normalization factors only,
\begin{equation}
\frac{P_V(I)}{P_I(V)} = \sqrt{\frac{\partial I^2} {\partial^2 {\cal E}}
\frac{\partial^2 {\cal E}}{\partial V^2}}.
\label{eq:proratio}
\end{equation}
The relations (\ref{eq:simplevolt}-\ref{eq:proratio}) are valid for the statistics of the outputs of the
compound circuit as well
as for the "black box" biased by either a voltage or a current source.
 
The FCS in the low-frequency limit
can  thus be  formulated in terms of stationary functions
${\cal E}_I,{\cal E}_V, {\cal E}(I,V)$.
These functions resemble thermodynamic  potentials. As those potentials, they are related
by Legendre transforms. However,
the relations presented concern  non-equilibrium systems.
They provide
an effective minimum principle for fluctuations
in such systems.

\subsection{Where are the Charging effects?}

 The electron transport through mesoscopic circuits is known
to be affected by charging or Coulomb blockade effects.
They  are due to  electron-electron interactions at the  mesoscopic scale.
The simplest case of a tunnel junction embedded into
an electromagnetic environment is described in \cite{Ingold}
in some detail.
Although these charging effects are not the topic of this
article, we find it important to note that they are present in
our formalism. Indeed, the Keldysh action
in use is  a generalization of the action
introduced in \cite{Schoen} to describe Coulomb blockade
effects in Josephson junctions.

How do charging effects manifest themselves in the low-frequency limit
considered?

The way to understand this is to invoke the field-theoretical
concept of renormalization. We have obtained the low-frequency
action  just by
substituting  slow realizations of
the fluctuating field $\vec{\chi}_2$ into the original action.
The renormalization procedure gives a more accurate way to obtain the
low-frequency action. In this procedure, the field is separated into
 a slow and a fast part. The path integration over the fast variables is
performed. Since fast and slow variables are coupled, this alters the functional of the slow variables, it is ``renormalized''. For the problem under consideration, this 
renormalization introduces  charging effects. Since
the environment action is  quadratic, it is not renormalized.
The renormalization can be thus ascribed to ${\cal E}_{\rm Sys}$.
The typical time scale
of relevant fast modes is $\tau_Q$. The importance of the charging
effects is governed by the value of the environment impedance at
the corresponding frequencies, $Z(1/\tau_Q)$\cite{Ingold}.  If $Z(1/\tau_Q) \ll
\hbar/e^2$, the renormalization is small and charging effects
can be disregarded. In the opposite case,  charging effects
are important and the renormalized action ${\cal E}_{\rm}$ differs
significantly from the bare one.

We thus conclude
that  charging effects
do not change the low-frequency relations given in the previous  subsection.
Their only effect is a renormalization of ${\cal E}_{\rm}$ in comparison
to its bare value at voltage bias. This renormalization  depends
on the environment response functions at the frequency scale $1/\tau_Q$.

\section{Mesoscopic conductor in a  macroscopic circuit}  \label{sc:QPC}

In this section, based on Ref.\ \cite{Kin02}, we illustrate the general relations obtained
above. We do this by considering a specific example:
a phase-coherent conductor in series with an impedance $Z$ (Fig. 2) 
\cite{Kin02}.
Further we focus on the shot noise limit $eV \gg k_BT$, and disregard
the Johnson-Nyquist noise produced by the linear resistor.
First we address the FCS of current fluctuations in the low-frequency
limit discussed above. We will thus make use of the relations
given in  section \ref{sc:low}. The assumption
$Z(\omega \simeq eV) \ll \hbar/e^2$ allows us to disregard
charging effects without  any restrictions on $Z(0)$.

To employ our scheme, we
first need the concrete expression for the FCS
of the voltage-biased conductor, ${\cal Z}_{\rm Sys}$.
In the  stationary case, this has been derived in \cite{Lev93}.
The mesoscopic conductor  is completely characterized by its transmission
eigenvalues $\Gamma_n$ \cite{report} that are assumed not to depend on
energy. In our notations, we have ($|eV| \gg k_B T$)

\begin{equation}
\label{eq:Zstatic}
{\cal Z}_{\rm Sys}(V,\chi)
= \exp \left\{ -{\cal E}_{\rm Sys} \tau \right\},\;\;
{\cal E}_{\rm Sys} =
-\frac{ e }{2\pi} |V| S[2ie
\chi\, {\rm sign}(V)]  \end{equation}
with the  dimensionless $S$ of  Eq.\ (\ref{eq:sofxi}).

Since we disregard the noise of the resistor, we have the environment action   ${\cal E}_{\rm Z}(V,\chi)=
-2i VZ^{-1}\chi$.
By virtue of Eq. (\ref{eq:simple}), the FCS of the circuit is defined
by
\begin{equation}
{\cal E}_I(V,\chi)= \mathop{{\rm extr}}_{V_2,\chi_2} \left\{
-2i Z^{-1}(V_2-V)(\chi_2-\chi) -\frac{ e |V_2|}{2\pi} S[2ie
\chi_2\, {\rm sign}(V_2)]  \right\}.
\end{equation}
Minimization with respect to $V_2$  fixes $\chi_2$,
\begin{eqnarray}
2 i Z^{-1}(\chi-\chi_2) = \frac{e}{2\pi} S(2ie\chi_2),
\label{eq:chifromchi2}
\end{eqnarray}
and makes  further minimization with respect to this
variable redundant.  We immediately arrive at
\begin{equation}
{\cal E}_I(V,\chi)= -\frac{e V}{2\pi} S[2ie\chi_2(\chi)]
\end{equation}
where $\chi_2(\chi)$ is implicitly defined  by Eq. (\ref{eq:chifromchi2}).
In the limit of a vanishing external impedance $Z \rightarrow 0$, we have 
$\chi_2 \to \chi$ so that ${\cal E}_I(V,\chi) \to
{\cal E}_{\rm Sys}$, as it should.

The coefficients of a series expansion of  ${\cal E}_I$ in $\chi$ give the cumulants of the
transmitted charge $\langle\!\langle Q^{p}\rangle\!\rangle$, whereas the coefficients of
${\cal E}_{\rm Sys}$ are the same cumulants $\langle\!\langle Q^{p}\rangle\!\rangle
_0$ in the limit  $Z\to 0$. Comparing the two series,
we obtain  relations between these cumulants.
The linear terms imply 
$\langle Q\rangle=(1+ZG)^{-1}\langle Q\rangle_{0}$,
$G$ being the conductance of the mesoscopic conductor.
This is a consequence of Ohm's law for
the average current $\bar{I}$ that  is
reduced by a factor $1+ZG$ by the resistor $Z$ in series.
The Langevin approach of Ref.\ \cite{Bla00} predicts  the same
re-scaling for the fluctuations of the current.
Indeed, in second order we find that
$\langle\!\langle Q^{2}\rangle\!\rangle=(1+Z_{2}G)^{-3}\langle\!\langle
Q^{2}\rangle\!\rangle_{0}$,  in agreement with this prediction.

However, for higher order cumulants we find  terms
that are not consistent with this re-scaling hypothesis.  For instance,
the third cumulant reads
\begin{equation}
\langle\!\langle Q^{3}\rangle\!\rangle=\frac{\langle\!\langle
Q^{3}\rangle\!\rangle_{0}} {(1+ZG)^{4}}- \frac{3ZG}
{(1+ZG)^{5}}\frac{\bigl(\langle\!\langle
Q^{2}\rangle\!\rangle_{0}\bigr)^{2}}{\langle
Q\rangle_{0}}.\label{thirdcumulant}
\end{equation}
Although the first term on the
 right-hand-side does have the scaling form conjectured,
the second term does not.
This is generic for  $p\geq 3$: $\langle\!\langle
Q^{p}\rangle\!\rangle=(1+ZG)^{-p-1}\langle\!\langle
Q^{p}\rangle\!\rangle_0$ plus a non-linear (multinomial) function of
lower cumulants. This mixing in of lower order cumulants is a consequence of the non-linear feedback mechanism discussed in the end of section  \ref{sc:pseudo}.
To give another example, the forth cumulant reads
\begin{eqnarray}
\langle\!\langle Q^{4}\rangle\!\rangle&=&\frac{\langle\!\langle
Q^{4}\rangle\!\rangle_{0}} {(1+Z_{2}G)^{5}} - \nonumber \\
&& - \frac{10 Z G}
{(1+Z_{2}G)^{6}}
\frac{\langle\!\langle
Q^{2}\rangle\!\rangle_{0} \langle\!\langle
Q^{3}\rangle\!\rangle_{0}}{\langle
Q\rangle_{0}}
+ \frac{15 Z^2 G^2}
{(1+Z_{2}G)^{7}}
\frac{\langle\!\langle
Q^{2}\rangle\!\rangle_{0}^3} {\langle
Q\rangle_{0}^2}. \label{forthcumulant}
\end{eqnarray}
We now turn to a
 one-channel conductor with transmission
probability $\Gamma$.
In this case, the distribution of
the integer transmitted  charge  $q\equiv I \tau e$ for voltage bias
$V$ is known to take the simple binomial form \cite{Lev93}
($\phi \equiv eV \tau /2\pi$)
\begin{equation}
P_{\phi}(q)={\phi \choose q}\Gamma^{q}(1-\Gamma)^{\phi-q}.
\label{Pqresult}
\end{equation}
The  distribution dual to this is that of the flux $\phi \equiv (e/h) \int_0^{\tau}{dt\,V(t)} $ that is accumulated at  the conductor under current
bias $I$.
We make use of  to the relation (\ref{eq:simplepseudo})
that determines these probabilities. For a mesoscopic conductor in the shot noise
limit, the corresponding ${\cal E}(I,V)$ reads
\begin{equation}
{\cal E}(I,V)= \mathop{{\rm extr}}_{\chi} \left\{
\frac{eV}{2\pi}S(2ie\chi) - 2i I \chi\right\}
\end{equation}
Expanding this near its point of extremum, we find
that
\begin{equation}
\frac{\partial I^2} {\partial^2 {\cal E}}
\frac{\partial^2 {\cal E}}{\partial V^2} = \frac{I^2}{V^2}
\end{equation}
for any functional form of $S$.
Employing  Eq. (\ref{eq:proratio}) we readily establish  that the
distribution function of flux $P_q(\phi)$ is given by (here we set
$q \equiv I\tau/e$)
\begin{equation}
P_{q}(\phi)= \frac{q}{\phi} P_{\phi}(q)=
{\phi-1 \choose q-1}\Gamma^{q}
(1-\Gamma)^{\phi-q},\label{Pphiresult}
\end{equation}
which is known  as the Pascal distribution.

We complement this derivation by a simple interpretation of the Pascal
distribution (\ref{Pphiresult}) of voltage fluctuations. 
The binomial distribution (\ref{Pqresult}) of transferred  charge 
can be interpreted \cite{Lev93} in gambling terms: it gives
the probability to win (transfer an electron)  $q$ times
in $\phi_0$ game slots given a winning chance of  $\Gamma$. In the voltage biased conductor, electrons leave the leads with a frequency $eV/h$ and  $\phi_0 = (e/h) \int_0^{\tau}{dt\,V(t)}$ is the number of electrons ("game slots") that try to pass the barrier in the detection time window.

The current bias changes the rules of the game. The gambling now ends 
when precisely $q_0$  attempts have been successful. This requires a fluctuating number $\phi$
of game slots (electrons that hit the barrier). Indeed, the Pascal distribution  quantifies the probability distribution of the number of independent
trials that one needs to achieve a given number of successes in a Bernoulli experiment.


\begin{thebibliography}{99}
\bibitem{Lev92}L. S. Levitov and G. B. Lesovik, JETP Lett. {\bf 55}, 555 (1992).
\bibitem{Lev93}L. S. Levitov and  G. B. Lesovik, JETP Lett. {\bf 58}, 230 (1993); L. S. Levitov, H. W. Lee, and G. B. Lesovik,
  Journ. Math. Phys. {\bf 37}, 4845 (1996).
\bibitem{Rammer}
  J. Rammer and H. Smith,
  Rev. Mod. Phys. {\bf 58}, 323 (1986).
\bibitem{Fey63}R.P. Feynman and F.L. Vernon, Ann. Phys.(N.Y.) {\bf
    24}, 118 (1963).
\bibitem{Naz99}Yu. V. Nazarov, 
  Ann. Phys. (Leipzig) {\bf 8} Spec. Issue SI-193, 507 (1999).
\bibitem{Bel01} W. Belzig and Yu. V. Nazarov, Phys. Rev. Lett. {\bf 87}, 197006 (2001).
\bibitem{She02}
A.\ Shelankov and  J.\ Rammer, cond-mat/0207343 (2002).
\bibitem{Naz01}
  Yu. V. Nazarov and M. Kindermann, cond-mat/0107133 (2001). 
\bibitem{Kin02}  M. Kindermann, Yu.V. Nazarov, and C.W.J. Beenakker,
cond-mat/0210617 (2002).
 \bibitem{Neumann} J. v. Neumann, {\em Die mathematischen Grundlagen der Quantenmechanik}, Springer Verlag, Berlin (1932).
\bibitem{CL} A.O. Caldeira and A.J. Leggett, Phys. Rev. Lett. {\bf 46}, 211 (1981).
\bibitem{Schoen} G. Sch\"{o}n and A.D. Zaikin, Phys. Rep. {\bf 198}, 237
(1990).
\bibitem{Alt00} A. Altland and A. Kamenev, Phys. Rev. Lett. {\rm 85}, 5615 (2000).
\bibitem{constraines} L. Kaplan, N. T. Maitra and E. J. Heller, Phys. Rev. A {\bf 56}, 2592 (1997).
\bibitem{Ingold} G.-L. Ingold and Yu. V. Nazarov, in {\it Single
Charge Tunneling}, edited by H. Grabert and M. H. Devoret,
NATO ASI Series B294 (Plenum, New York, 1992).
\bibitem{report}
  C. W. J. Beenakker, Rev. Mod. Phys. {\bf 69}, 731 (1997).

\bibitem{Bla00} Ya.\ M. Blanter and M. B\"{u}ttiker, Phys.\ Rep.\ {\bf 336}, 1
(2000). The effect of a series resistance on the noise power is discussed in
\S~2.5.
\end{thebibliography}
\end{document}